\documentclass[aps,pra,reprint,10pt,a4paper]{revtex4-1}
\usepackage[utf8]{inputenc}
\usepackage[sc,osf]{mathpazo}
\usepackage{amsmath}
\usepackage[T1]{fontenc}
\usepackage{latexsym}
\usepackage{amssymb}
\usepackage[colorlinks=true,citecolor=blue,urlcolor=blue]{hyperref}
\usepackage{color}
\usepackage{graphics,epstopdf}
\usepackage{soul}
\usepackage[demo]{graphicx}
\usepackage{capt-of}
\usepackage{lipsum}
\usepackage{adjustbox}
\usepackage[normalem]{ulem}
\usepackage[table,xcdraw]{xcolor}
\usepackage{braket}
\usepackage{physics}
\usepackage[normalem]{ulem}
\usepackage{soul}

\begin{document}

\title{
Multimode advantage in continuous variable quantum battery}

\author{Tanoy Kanti Konar$^1$, Ayan Patra$^1$, Rivu Gupta$^1$, Srijon Ghosh$^{1,2}$, Aditi Sen(De)$^1$}

\affiliation{$^1$ Harish-Chandra Research Institute,  A CI of Homi Bhabha National
Institute, Chhatnag Road, Jhunsi, Prayagraj - 211019, India \\
$^2$ Faculty of Physics, University of Warsaw, Pasteura 5, 02-093 Warszawa, Poland}

\begin{abstract}

We provide an architecture for a multimode quantum battery (QB) based on the framework of continuous variable (CV) systems. We examine the performance of the battery by using a generic class of multimode initial states whose parameters can be tuned to produce separable as well as entangled states and that can be charged locally as well as globally by Gaussian unitary operations. Analytical calculations show that a separable state is equally advantageous to an entangled one for two- and three-mode batteries when taking the figures of merit as the second moments of the change in energy.  In order to produce a stable quantum battery consisting of an arbitrary number of modes, we derive compact analytical forms of the energy fluctuations and prove that for a multimode separable Gaussian initial state, fluctuations decrease as the number of modes increases, thereby obtaining a scaling analysis. Moreover, we demonstrate that local displacement as a charger is better for minimizing the fluctuations in energy than that involving the squeezing unitary operation.

\end{abstract}

\maketitle

\section{Introduction}
\label{sec:intro}

Small-scale energy-storing devices like batteries are necessary due to the growing trend of miniaturizing electronic devices, especially at the molecular and subatomic levels. Batteries typically store chemical energy and convert it into electrical energy through well-known reduction-oxidation reactions. 
Although these batteries play a crucial role in our daily lives, starting from medical instruments to household appliances, navigation, and also in transportation systems, they have several significant constraints at the microscopic level. 
The principles of quantum mechanics have been shown to modify the laws of thermodynamics \cite{breuer2002,rivas2012}  and improve the performance of a variety of devices, including that of quantum thermal machines \cite{popescu10,joulain16}.

A quantum battery (QB) is a device that can store energy, and from which the energy can be extracted via unitary operations. It typically consists of a collection of $d$-dimensional quantum systems governed by a Hamiltonian having a non-degenerate energy spectrum \cite{Alicki}. 
In the last few years, a series of QB models, charged via global and local unitary operators, have been proposed, which include many-body quantum batteries \cite{Modispinchain,andolina2017, campaioli2017, andolina2019, santos2019, santos2020}, interacting spin-chains \cite{srijon2020}, Dicke quantum battery \cite{ alba_1_2020, dou2022}, quantum battery based on superconducting qubits \cite{hu2021optimal} and batteries formulated with the Sachdev-Ye-Kitaev (SYK) \cite{andolina2020} and also the Lipkin-Meshkov-Glick (LMG) models \cite{farre2020,abah2022,dou2022_1}. 
Numerous investigations have also been conducted to analyze the detrimental impact on the performance of the QB by taking into account the inevitable interaction of the system with the environment \cite{Giovannetti2019, barra2019, alba_2_20, zhao2021, srijon21}.
 Going beyond these distinguishable systems, the modeling of energy-storing devices with indistinguishable particles, with ultra-cold atomic setups in the test bed of bosons and fermions, has also been addressed in recent times \cite{konar2022}. Despite the tremendous progress in the field of quantum batteries, it is still unclear how non-classicality, like quantum entanglement\cite{HHH2009} or quantum coherence\cite{baumgratz2014} affects the performance of quantum thermal machines. In particular, global entangling operations are found to be more beneficial than the local operations, which are further supported by a very recent work \cite{gyhm2022}, although it may not generate entanglement between the subsystems \cite{hovhannisyan2013, Binder2015, arjmandi2022}.


All the aforementioned QBs are designed on systems having discrete and finite degrees of freedom. Interesting studies on QBs in infinite dimensional systems are limited. Variables like the position and momentum of a particle possess a continuous spectrum and are used to characterize continuous variable (CV) systems. Prominent examples include the harmonic oscillator, through which many physical systems having an important role in quantum optics \cite{Serafini_2017} have been realized and which also provide significant benefits in the fields of quantum error correction, and several quantum information processing tasks \cite{braunstein1998_1, braunstein1998_2, braunstein1998, lloyd1999, ralph1999, li2002, mizuno2005, furusawa1998,bowen2003, andersen2005, lance2005}.
As a thermal machine, 
harmonic oscillators are also used as a charger to deposit energy in a two-level system \cite{andolina2018,yong2021}, while non-Gaussian charging is shown to be optimal \cite{friis2018} to charge a QB made of infinite-dimensional bosonic systems 
under the constraint of low environmental temperature and in the presence of strong squeezing \cite{centrone2021}. 
All these proposals for CV batteries involve few or a single mode. 
\begin{figure*}
    \centering
    \includegraphics[scale=0.24]{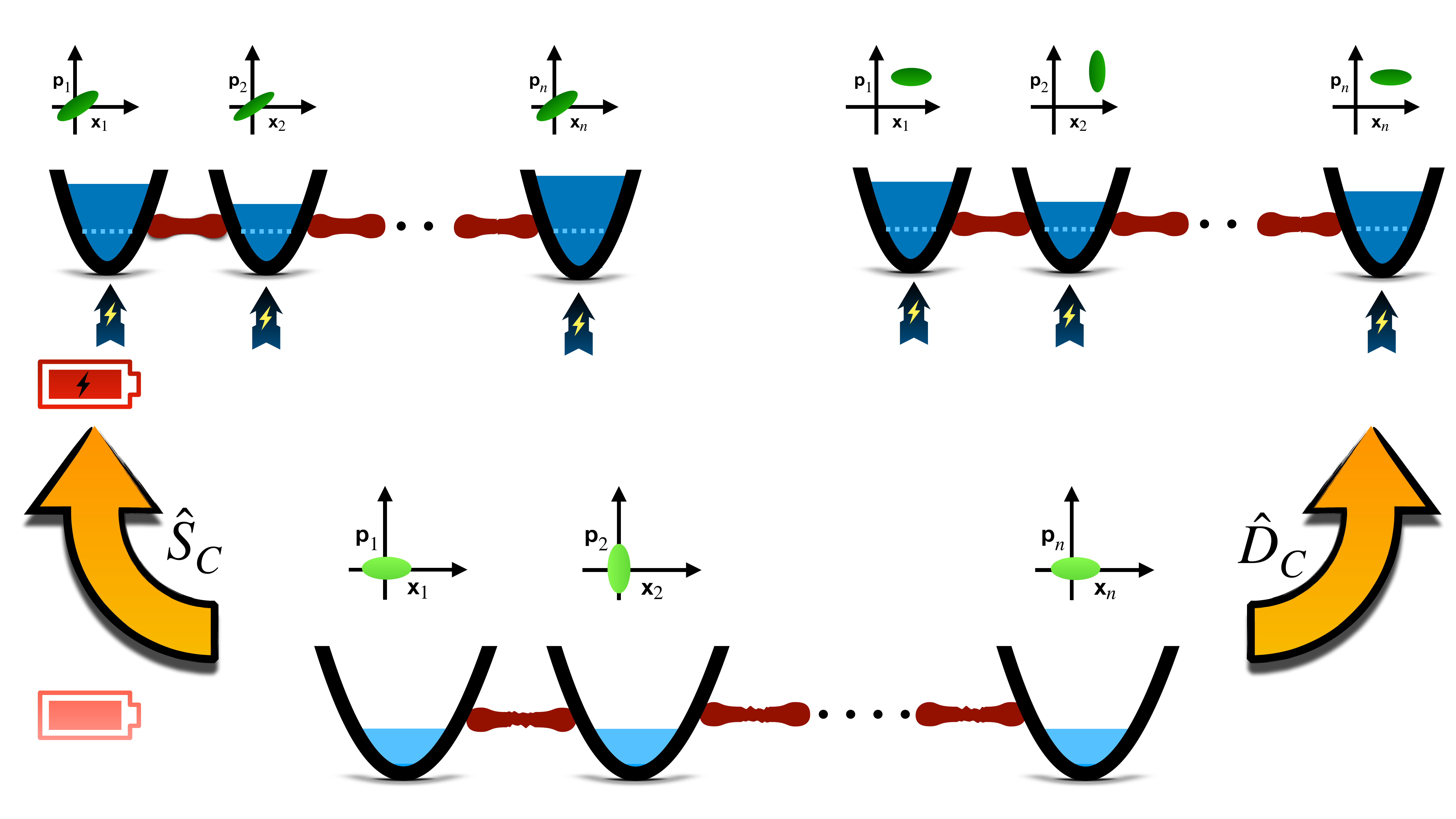}
    \caption{Schematic diagram depicting the functioning of a multimode continuous variable quantum battery. Initially, it is uncharged and has a minimum amount of energy in each mode which is shown as filled in light (blue) color. The phase space description of each mode is taken to be the same and is shown as light (green) ellipses. The local charging operations are considered to be either displacement (\(\hat{D}_c\)) or squeezing (\(\hat{S}_c\)). The dark (blue) color represents the increment in the energy of the charged battery from the dotted line. The charged modes are either squeezed or displaced which are depicted as dark (green) ellipses in phase space.}
    \label{fig:cv_schematics}
\end{figure*}

In this paper, we go beyond this restriction and explore the question -- {\it does the battery's efficiency increase if it is constructed with several modes? If so, is multimode entanglement required for this improvement?}
We report here that although the second answer is negative, the first one is affirmative. \textcolor{black}{Specifically, we present a design of a QB composed of an arbitrary number of non-interacting Gaussian systems. These systems are charged using local Gaussian operations, such as squeezing and displacement \cite{Brown_2016}, to demonstrate a quantum advantage in terms of the increasing number of modes (see Fig. \ref{fig:cv_schematics} for a schematic representation of our model).} To analyze the performance,  the second moments in the energy change, known as charging precision \cite{friis2018} -  determining the stability of the battery, and the work fluctuation - quantifying how accurately the battery can be charged, are computed after optimizing the parameters involved in the charging process and the initial state of the battery. Specifically, we choose  
a class of generic states comprising two- and three-modes with varying degrees of entanglement, as the initial states of the QB, thereby representing both separable and entangled states. 
We highlight that when charging is carried out using global and local operations, and the energy to be stored is fixed to a given amount, fully separable and entangled states provide an identical level of stability in the QB. \textcolor{black}{Note here that our findings counter the established result that applying global unitary operations generally reduces work precision \cite{llobet_njp_2019, Bakhshinezhad_PRE_2024}. This deviation arises partly because we focus on a restricted set of operations and initial states that, while experimentally feasible, yield these counterintuitive outcomes. Furthermore, our investigation is strictly limited to the Gaussian regime, and thus, we do not explore the potential advantages of entanglement in the non-Gaussian domain. It should also be noted that implementing non-Gaussian operations is highly probabilistic in practice.} Our results demonstrate that the modeling proposed here definitely lowers the experimental cost of the CV battery.


By charging the system with local squeezing and displacement operations, we also demonstrate the multimode advantage of the stability of the CV battery. In particular, our analyses, supplemented with numerical data, show that by considering initial separable states with an arbitrary number of modes, the fluctuation during discharging decreases sharply with the increase of modes, when squeezing unitaries are used as the charging operation. Although the work fluctuation does not vary with the number of modes in the case of local displacement, the fact that the charging precision decreases as the number of modes grows, once more emphasizes the constructive role played by the multiple modes present in the battery. The minimum values attained by both the figures of merit are lower in the case of local displacement as charging operations than that provided through squeezing. 
Such modal superiority has a two-fold implication: firstly, it boosts the stability of the battery against fluctuations in energy, and secondly,  storing more energy in the battery is possible, since charging a single mode is restricted by experimental limitations (especially in the case of squeezing). Therefore, a higher number of modes directly implies a more perfectly functioning CV quantum battery.\\\\
The paper is organized as follows: In Sec. \ref{sec:battery_fig}, we introduce the basic notions of CV quantum battery and its charging operations along with the figures of merit. The importance of considering the entangled state as the initial state and the role of global as well as local charging are reported in Sec. \ref{sec:ent_need} for effectively small system size. Motivated by the results obtained in Sec. \ref{sec:ent_need}, a scaling analysis is performed by considering a fully separable initial state consisting of an arbitrary number of modes, and  the multimodal advantage is reported in Sec. \ref{sec:N-mode_sep}. Finally, in Sec. \ref{sec:conclu}, we summarize our results.

\section{Continuous variable quantum battery and its figures of merit}
\label{sec:battery_fig}

A quantum battery consists of an arbitrary number of quantum systems, in which energy can be stored and extracted via quantum operations. The internal energy is described by the battery Hamiltonian, $\hat{H}_B$. To store (charging) or extract (discharging) work from the system, the battery is evolved by a charging Hamiltonian, $\hat{H}_{C}$, which must be non-commuting with respect to \(\hat{H}_\text{B}\). The choice of the battery Hamiltonian, and the charging Hamiltonian depends on the system of interest \cite{andolina2017,campaioli2017,Giovannetti2019,santos2019,srijon2020,zhao2021}.


\subsubsection{ Design of the CV Battery} 
\label{subsubsec:battery_formalism}
Instead of working in finite dimensions, we propose a quantum battery built of infinite dimensional systems, specifically continuous variable (CV) systems. We choose the initial state of the battery as an \(\mathcal{N}\)-mode Gaussian state which can be described by its first and second moments in the phase space. The underlying Hamiltonian describing the energy of the system is $\hat{H}_B=\sum_{j=1}^{\mathcal{N}}\omega_j\hat{N}_j$, where \(\hat{N}_j=\hat{a}_j^\dagger \hat{a}_j\) denotes the number operator. In other words, we consider the system consisting of \(\mathcal{N}\) harmonic oscillators, each of frequency \(\omega_j\) in the mode $j$. We choose the initial state as both correlated and uncorrelated, which, e.g. can be prepared through the application of beam splitters on an \(\mathcal{N}\)-mode product state with alternate squeezing \(r, -r, ...\) in consecutive modes \cite{Patra_arxiv_2022}. Hence, the class of correlated states can be characterized by $\mathcal{N}-1$ parameters family, $\{\tau_j\}_{j=1,2,...,(\mathcal{N}-1)}$ where $\tau_j$ represent the transmittivity of the $j$-th beam splitter. Note that most of the well-known entangled states can be constructed in this way as we will describe later. 

\textcolor{black}{{\it Battery operation.} The objective of the charging operation is to augment the energy of the system by considering the work stored, particularly in the case of closed dynamics, whereas the discharging operation is intended to extract the stored energy.} This evolution changes the state of the system to another state which, in turn, creates an energy difference \(\Delta E_{U_C}\), where \(U_C\) denotes the charging unitary. We analyze the stability of the battery during both charging and discharging. We also provide the most efficient modal distribution of the energy, \(\Delta E_{{U_C^j}}\), in an arbitrary mode $j$, such that \(\sum_j\Delta E_{{U_C^j}} = \Delta E_{U_C}\). As charging operations, local squeezing and displacement unitaries act on each mode while global squeezing operations are performed on the entire system. Both pictures address the question of whether entanglement in the initial state or entangling charging operations can help to achieve optimal figures of merit. 
	\begin{itemize}
	    \item {\bf Squeezing operation.} The squeezing operator acting on the mode \(j\) is given by  $\hat{S}_j(\zeta_j) = \exp \Big[\frac{1}{2}(\zeta_j \hat{a}_j^{\dagger^2} - \zeta_j^* \hat{a}_j^2)\Big]$ where, $\zeta_j = \delta_j e^{i \theta_j}$ is the squeezing parameter, with $\delta_j$ and $\theta_j$ respectively being the squeezing degree and the squeezing angle. Here, $i = \sqrt{-1}$ and $*$ denotes complex conjugation. The total local squeezing operator for the $\mathcal{N}$-mode battery is given by  $\hat{S}_c = \hat{S}_1(\zeta_1)\otimes\ldots\otimes\hat{S}_{\mathcal{N}}(\zeta_{\mathcal{N}})$. We refer to $\theta_j$ as the phase of the charging operation.

            \item {\bf Displacement operation.} If the battery is charged via local displacements, the charging operator on the \(\mathcal{N}\) modes reads \(\hat{D}_c = \hat{D}_1(\alpha_1) \otimes\ldots\otimes \hat {D}_{\mathcal{N}}(\alpha_{\mathcal{N}})\) where $\hat{D}_j (\alpha_j) = \exp \Big[ \alpha_j \hat{a}_j^{\dagger} - \alpha_j^* \hat{a}_j\Big]$ is the local displacement operator characterized by  $\alpha_j = |\alpha_j| e^{i \phi_j}$ with $\phi_j$ being the phase of the charging unitary.
	
	\end{itemize}
	
{\it Performance indicators. } Instead of studying the usual work output of the battery, which is typically studied in spin systems \cite{Batteryreview}, we are interested in quantifying the stability of the quantum battery. Precisely, the charging process can be designed in such a way, that at the time of discharge, it can produce the same amount of energy that is stored during charging. This is quantified by a parameter known as the charging precision \cite{friis2018},  which is defined for an \(\mathcal{N}\)-mode system as
	\begin{equation}
		\Delta \sigma^{(\mathcal{N} )}=\sqrt{V({\rho_1})}-\sqrt{V({\rho_0})},
            \label{eq:sigma_defn}
	\end{equation}
	where \(V({\rho_k})\) stands for the variance, given by 
 \begin{equation}
	    V({\rho_k}) = \tr[\hat{H}_B^2\rho_k]-\big(\tr[\hat{H}_B\rho_k]\big)^2.
     \label{eq:battery_V}
	\end{equation} 
Here, \(\rho_k\) with $k = 0,1$ corresponds to the initial and final states of the battery respectively, and \(\rho_1=U_\text{C}\rho_0 U_\text{C}^\dagger\). \\ 
 If the charging strength is $\gamma_j \in \{\delta_{j}, |\alpha_{j}| \}$ and its corresponding phase is $\nu_j \in \{\theta_{j}, \phi_{j}\}$ for $j$-th mode of the battery, we have\textcolor{black}{
 \begin{equation}
     \Delta \sigma^{(\mathcal{N})} = f_1(s,\{\omega_j\}, \{\gamma_j\}, \{\nu_j\}),
     \label{eq:Dsigma_opt_1}
 \end{equation}
for some function $f_1$. Here $s$ represents multiparameters, consisting of the initial state parameters e.g., the initial squeezing $r$ and beam splitter transmittivities $\tau_i$s (i.e., $s=\{r,\tau_1,...,\tau_{\mathcal{N}-1}\}$), and \(\omega_j\)s are the parameters of the battery Hamiltonian. Note that $\{x_j\}$ is a collection of $\mathcal{N}$ elements, 
i.e., $\{x_j\}=\{x_1,...,x_{\mathcal{N}}\}$ with $x$ being $\omega, \gamma,$ or $\nu$.} For the mode $j$, we can invert the expression for a fixed energy increment $\Delta E_{{U_C^j}}$ to obtain $\gamma_j$, i.e.,
\begin{eqnarray}
   \nonumber \Delta E_{{U_C^j}} &=& f_2(s,\omega_j, \gamma_j, \nu_j) \text{\textcolor{black}{~for an invertible function $f_2$,}}\\
   \implies \gamma_j &=& f_3 (s,\omega_j, \Delta E_{{U_C^j}}, \nu_j). \label{eq:Dsigma_opt_2}
\end{eqnarray}
Using Eq. \eqref{eq:Dsigma_opt_2}, we can obtain the charging precision as a function of the increased  energy of the different modes, i.e.,
\begin{equation}
  \Delta \sigma^{(\mathcal{N})}=f_U^{(\mathcal{N})} \left(s,\{\omega_j\},\left\{\Delta E_{U_C^j}\right\},\{\nu_j\}\right),
    \label{eq:Dsigma_opt_3}
\end{equation}
with $U$ being $S$ (or, $D$) for squeezing (or, displacement) operation. By fixing the initial state parameters $s$, and the total energy increment, $\Delta E_{U_C} = \sum_{j = 1}^{\mathcal{N}} \Delta E_{U_C^j}$, our task is to minimize  $\Delta \sigma^{(N)}$, over the set of the charging phases $\{\nu_j\}$, and the energy distribution, $\{\Delta E_{U_C^j}\}$. This minimum value is denoted as $\Delta \sigma_{\min}^{(\mathcal{N})}$. Although a general battery Hamiltonian includes arbitrary \(\omega_j\)s, for ease of calculation, we assume that they are all equal to unity. \textcolor{black}{Consequently, the average energy and mean square energy for the $j$-th mode are given by $E_{U_C^j}=\langle \hat{N}_{U_C^j} \rangle$ and $E^2_{U_C^j}=\langle \hat{N}^2_{U_C^j} \rangle$, respectively, where $\hat{N}_{U_C^j}$ is the number operator for mode $j$ after the application of $U_C^j$}. However, the conclusions remain the same even under this assumption, since the overall behavior is qualitatively similar when one considers arbitrary \(\omega_j\)s. In particular, the difference in taking arbitrary \(\omega_j\)s would be reflected in the optimal energy distribution $\{\Delta E_{{U}_C^j}\}$ of the charging process and the optimal charging phases $\{\nu_j\}$, although the effect of the number of modes \textcolor{black}{as well as the existing correlations among them} on the figures of merit remains unaffected.

 Another indicator of the performance of the battery is the energy fluctuation during the charging process. Towards defining it, let us first rewrite \(\Delta E_{U_C}\) as 
	\begin{eqnarray}
	\nonumber\Delta E_{U_C}&&=\tr[\hat{H}_B\rho_1]-\tr[\hat{H}_B\rho_0]\nonumber\\&&= \tr[\hat{H}_B U_{C}\rho_0U_{C}^{\dagger}]-\tr[\hat{H}_B\rho_0]\nonumber\\&&=\tr[(\hat{H}_B'-\hat{H}_B)\rho_0]\nonumber\\&&=\tr[(\Delta \hat{H}_B)\rho_0],
 \label{eq:delta_E_defn}
	\end{eqnarray}
	where  \(\hat{H}_B' = U_{C}^{\dagger}\hat{H}_B U_{C}\). To store \(\Delta E_{U_C}\) amount of energy in the battery by the corresponding charging process \(U_C\), the deviation of stored energy for an \(\mathcal{N}\)-mode system is quantified by a quantity called work fluctuation \cite{friis2018}, which is defined as
	\begin{eqnarray}
		\nonumber \Delta W^{(\mathcal{N})}&&=\nonumber\sqrt{ \tr[\Delta \hat{H}_B^2\rho_0]-\big(\tr[\Delta \hat{H}_B\rho_0]^2\big)}\nonumber\\&&=\sqrt{V(\rho_1)+V(\rho_0) - 2~ \text{Cov}(\hat{H}_B',\hat{H}_B)}, ~~~~~~
  \label{eq:W_defn}
	\end{eqnarray}
	where \(\text{Cov}(\hat{H}_B',\hat{H}_B)=\frac{1}{2}\expval{\{\hat{H}_B',\hat{H}_B\}}_{\rho_0}-\expval{\hat{H}_B'}_{\rho_0}\expval{\hat{H}_B}_{\rho_0}\) and the expectation value is taken with respect to the  initial state \(\rho_0\). Note that $\expval{\hat{H}_B}_{\rho_0}=E_0^{(\mathcal{N})}$ and $\expval{\hat{H}_B'}_{\rho_0}=E_1^{(\mathcal{N})}$ are the energies of the uncharged and charged states of the battery respectively, which give $\Delta E_{U_C} = E_1^{(\mathcal{N})} - E_0^{(\mathcal{N})}$. We will use \(\Delta W^{(\mathcal{N})}\) as an indicator to check the stability of the QB during the charging process. The optimization of the work fluctuation $\Delta W^{(\mathcal{N})}$ proceeds along the same line as that of $\Delta \sigma^{(\mathcal{N})}$ as elucidated through Eqs \eqref{eq:Dsigma_opt_2} - \eqref{eq:Dsigma_opt_3}, \textcolor{black}{where the work fluctuation can be obtained as $\Delta W^{(\mathcal{N})} = g_U^{(\mathcal{N})} \left(s,\{\omega_j\},\left\{\Delta E_{U_C^j}\right\},\{\nu_j\}\right)$ for some function $g_U$ with $\omega_j$s being unity as mentioned beforehand. For a given initial state parameters, $s$, and total energy increment, $\Delta E_{U_C}$, the minimum work fluctuation will be denoted as $\Delta W_{\min}^{(\mathcal{N})}$.}

    For a well-functioning battery, we are required to minimize both the energy fluctuation and charging precision, to make the battery stable during the charging and discharging processes which, in turn, implies the minimization of both \(\Delta \sigma^{(\mathcal{N})}\) and \(\Delta W^{(\mathcal{N})}\), thereby achieving $\Delta \sigma_{\min}^{(\mathcal{N})}$ and $\Delta W_{\min}^{(\mathcal{N})}$. \textcolor{black}{In general, these quantities are not identical at finite temperatures and also can not be optimized simultaneously \cite{friis2018,crescente_njp_2020}. In this context, we aim to optimize work fluctuation during the charging process, achieving $\Delta W_{\min}^{(\mathcal{N})}$. On the other hand, we focus on optimizing charging precision during the discharging phase, leading to $\Delta \sigma_{\min}^{(\mathcal{N})}$.  
    Hence, our objective is to increase the energy of the battery by an amount \(\Delta E_{U_C}\), which is distributed among the different modes, and check whether it is helpful to increase the number of modes of the QB and to estimate the $\Delta \sigma_{\min}^{(\mathcal{N})}$ and $\Delta W_{\min}^{(\mathcal{N})}$ independently. } 

\section{Necessity of entanglement for multimode batteries}
\label{sec:ent_need}
In order to analyze the performance of the battery consisting of \(\mathcal{N}\)-modes, it is natural to investigate whether entanglement is necessary for the proper functioning of a multimode quantum battery. 
\textcolor{black}{Unlike several finite dimensional systems \cite{hovhannisyan2013}, we report that entanglement does not provide any additional benefit in the set up considered here.}
 Note that both entangled and product states can be prepared from the ground state of the battery Hamiltonian, by global and local operations respectively. Specifically, we compare the trends of $\Delta \sigma^{(\mathcal{N})}$ and $\Delta W^{(\mathcal{N})}$ obtained from two- and three-mode entangled states with those from squeezed separable states as initial states. 
\subsection{Two-mode CV battery: Entangled vs. Product states}
\label{subsec:2-mode-entangled}
To design the initial state of a two-mode battery, we use a beam splitter $BS_{12}$ with transmittivity $\tau$ where $0\leq\tau\leq1$. Two squeezed vacuum modes, one squeezed in the position quadrature and the other squeezed in the momentum quadrature with the same strength, $r$, are impinged on the beam splitter, and an entangled output state, characterized by $\tau$ and $r$ is generated. Though the transmittivity (i.e., the state parameter) can take its value upto $1$, the entanglement content of the output state is symmetric with $\tau = 0.5$. It can be shown that the entanglement of the output state increases with \(\tau\) upto \(\tau=0.5\) which leads to the well-known two-mode squeezed vacuum (TMSV) state, the maximally entangled state for a given energy, while $\tau = 0$ and $1$ correspond to a separable state. \textcolor{black}{The representation of the state in the phase-space formalism is provided in Appendix. \ref{app:states}.}

Note that initially, the energy of the uncharged battery is \(E_0^{(2)} = 2 \sinh^2r\). Such a choice of the initial state can help us to probe entangled states having different entanglement content, as well as the product state. Squeezing and displacement operations on each mode and the global two-mode squeezing operator are applied to store energy in the battery. In each case, we aim to determine the optimal value of $\tau$, and, therefore, the entanglement, for which charging in both the modes yields the minimum value of $\Delta \sigma^{(2)}$ and $\Delta W^{(2)}$.
\subsubsection{Charging with local squeezing}
\label{subsubsec:squeezing_2-mode_tau}
\begin{figure*}
    \centering
    \includegraphics[width=1.00 \linewidth]{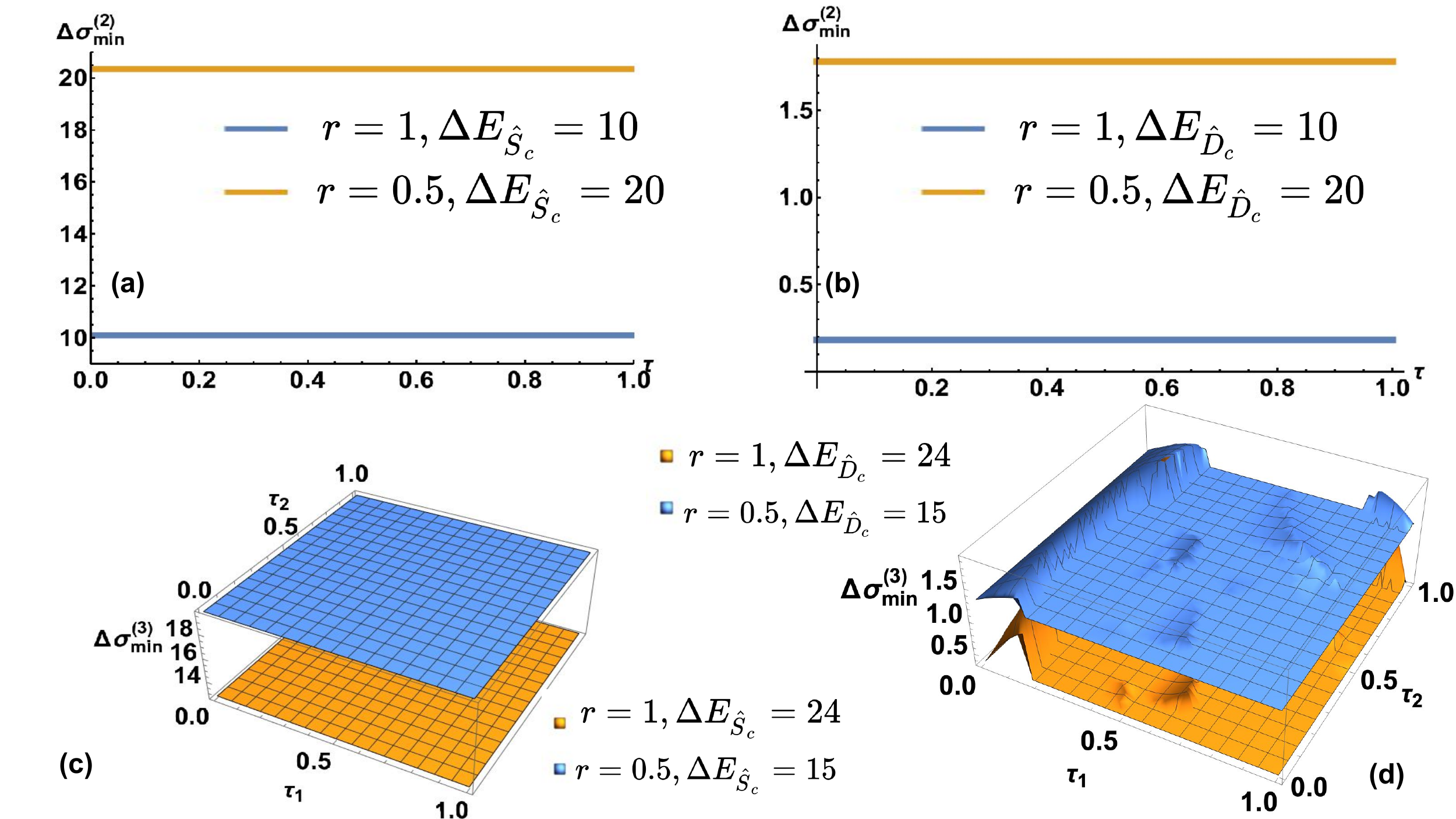}
    \caption{(Color Online.) \textbf{Behavior of the optimal charging precision, $\Delta \sigma_{\min}^{(\mathcal{N})}$ for two- and three-mode entangled initial states}.  The trends of $\Delta \sigma_{\min}^{(2)}$ (ordinate) are depicted by varying transmittivity $\tau$ (abscissa) of the beam splitter   for charging through squeezing and displacement unitaries in (a) and (b) respectively.  The net energy increase of the battery is $\Delta E_{\hat{D}_c} (\Delta E_{\hat{S}_c}) = 10$ for $r = 1$ (dark blue) and $\Delta E_{\hat{D}_c} (\Delta E_{\hat{S}_c}) = 20$ for $r = 0.5$ (light orange). In (c) and (d),  $\Delta \sigma_{\min}^{(3)}$ (z-axis) is shown against $\tau_1$ (x-axis) and $\tau_2$ (y-axis) for squeezing and displacement chargers of the three-mode initial states respectively. The energy specifications in this case correspond to $r = 0.5, \, \Delta E_{\hat{D}_c} (\Delta E_{\hat{S}_c}) = 15$ (dark blue) and $r = 1, \, \Delta E_{\hat{D}_c} (\Delta E_{\hat{S}_c}) = 24$ (light orange). All the axes are dimensionless.}
    \label{fig:2-3_modes}
\end{figure*}

The charging operator for the two-mode battery consists of local squeezing operators acting on the individual modes, given by  $\hat{S}_c = \hat{S}_1(\zeta_1) \otimes \hat{S}_2(\zeta_2)$, where the local squeezing operators $\hat{S}_j(\zeta_j)$ have been defined in Sec. \ref{subsubsec:battery_formalism}. Upon charging, the increase in the total energy of the battery is $\Delta E_{\hat{S}_c} = \sum_{j = 1}^2 \Delta E_{\hat{S}_c^j}$, where

\begin{eqnarray}
    \nonumber  \Delta E_{\hat{S}_c^1} &=& \sinh \delta_1 (\sinh \delta_1 \cosh 2 r\\&&\nonumber+(2 \tau-1) \cosh \delta_1 \cos \theta_1 \sinh 2 r), \\
    \text{and}\quad
     \nonumber \Delta E_{\hat{S}_c^2} &=& \sinh \delta_2 (\sinh \delta_2 \cosh 2 r\\&&+(1 - 2 \tau) \cosh \delta_2 \cos \theta_2 \sinh 2 r),
    \label{eq:deltaE1_2-mode_tau_sq}
\end{eqnarray}
represent the increase in energy of the two modes. We can similarly calculate the second moments of the energy and hence can obtain the charging precision $\Delta \sigma^{(2)} = f_S^{(2)}(r, \Delta E_{\hat{S}_c^j}, \theta_j, \tau)$ for a fixed squeezing strength of the initial state, \(r\).
We distribute the total energy $\Delta E_{\hat{S}_c}$ as $k \Delta E_{\hat{S}_c}$ and $(1 - k)\Delta E_{\hat{S}_c}$ in the two modes and find that the charging precision is minimum at $k = 1/2$ after optimizing over \(k, \theta_j\), and \(\tau\).
Now, while finding the optimum entanglement content of the battery, i.e., the optimal value of $\tau$
for a given $\Delta E_{\hat{S}_c}$, at $\Delta E_{\hat{S}_c^1} = \Delta E_{\hat{S}_c^2} = \Delta E_{\hat{S}_c}/2$,  $\Delta \sigma^{(2)}_{\min}$ turns out to be independent of \(\tau\) \textcolor{black}{(and hence independent of the entanglement content)}
 such that $\theta_1+\theta_2=\pi$, for a fixed $r$, 
 as depicted in Fig. \ref{fig:2-3_modes}(a).
  This is true irrespective of the initial squeezing strength (initial energy of the uncharged battery) and the energy provided to the battery via the charging process.\\

\textcolor{black} {In order to calculate the work fluctuation, $\Delta W^{(2)}$, as described in Eq. (\ref{eq:W_defn}), we need to compute the variance of $\hat{H}_B$ in the states $\rho_0$ and $\rho_1$ and the covariance term $\text{Cov}(\hat{H}_B',\hat{H}_B)$ in the state $\rho_0$. Henceforth, our main task is to calculate $\langle\{\hat{H}_B',\hat{H}_B\}\rangle_{\rho_0}$. For charging with local squeezing, $\hat{H}_B'= \hat{S}_c^\dagger\hat{H}_B\hat{S}_c$. Using the relation $\hat{S}_j^\dagger(\zeta_j)\hat{a}_j\hat{S}_j(\zeta_j)=\hat{a}\cosh \delta_j + \hat{a}_j^\dagger e^{i\theta_j}\sinh\delta_j$, we obtain
 \begin{equation*}
     \hat{H}_B'=\sum_{j=1,2}\left(\hat{N}_j\cosh 2\delta_j + \hat{A}_j\sinh 2\delta_j + \sinh^2\delta_j\right),
 \end{equation*}
 with $\hat{A}_j=(\hat{a}_j^2 e^{-i \theta_j} + \hat{a}_j^{\dagger 2} e^{i \theta_j})/{2}$. Leveraging the Wigner function formalism (see Appendix \ref{sec:CV}), we finally obtain 
 \begin{widetext}
  \begin{eqnarray*}
     \langle\{\hat{H}_B',\hat{H}_B\}\rangle_{\rho_0}=&&2\left(\left(1+2\cosh 2r\right)\left(\cosh 2\delta_1+\cosh 2\delta_2\right)-2\right)\sinh^2 r \\&&+ \left(1-2\tau\right)\left(\sinh 2r - \sinh 4r\right)\left(\cos\theta_1 \sinh 2\delta_1 - \cos\theta_2 \sinh 2\delta_2\right).
 \end{eqnarray*}
 \end{widetext}
 Substituting the value of $\langle\{\hat{H}_B',\hat{H}_B\}\rangle_{\rho_0}$ and $\langle \hat{H}_B\rangle_{\rho_{i=0,1}}$ in the expression of $\text{Cov}(\hat{H}_B',\hat{H}_B)$, and using Eq. (\ref{eq:deltaE1_2-mode_tau_sq}), we obtain $\Delta W^{(2)}=g_S^{(2)}(r, \tau, \Delta E_{\hat{S}_c^j}, \theta_j)$, for some function $g_S^{(2)}$. If we fix the initial squeezing, $r$, and total energy increment, $\Delta E_{\hat{S}_c}$, for a given $\tau$, the minimization of $\Delta W^{(2)}$ is achieved at the optimal phase $\theta_1=\pi (\text{or}, 0)$ and $\theta_2=0 (\text{or}, \pi)$ for $0\leq\tau\leq1$ with equal energy distribution. The variation of $\Delta W^{(2)}_{\min}$ with respect to the state parameter $\tau$ is illustrated in Fig. \ref{fig:2-3_modesW} (a), for different values of \(r\) and $\Delta E_{\hat{S}_c}$. Noticeably, $\Delta W^{(2)}_{\min}$  increases with entanglement, highlighting the detrimental impact of entanglement.
 }

\textbf{Remark 1}. \textcolor{black}{For the fully separable state ($\tau = 0$), when charged via squeezing operation, we find that the value of $\Delta \sigma_{\min}^{(2)}$ is independent of $\theta_j$.} This indicates that, for a two-mode separable battery, the optimal stability is solely determined by the squeezing strength of the charging operation in each mode, regardless of the values of \(\theta_1\) and \(\theta_2\). In contrast, for an entangled system used as a battery, $\Delta \sigma_{\min}^{(2)}$ is a function of \(\theta_1\) and \(\theta_2\). This suggests that separable states provide greater flexibility in selecting the optimal configuration compared to entangled states.


\subsubsection{Charging via displacement}
\label{subsubsec:disp_2-mode_tau}
Instead of squeezing unitary, let us find whether a change of unitary, i.e., when the battery is charged via local displacements in each mode can lead to a more stable QB or not. In particular, the charging operator reads $\hat{D}_c = \hat{D}_1(\alpha_1) \otimes \hat {D}_2(\alpha_2)$, where $\hat{D}_j (\alpha_j)$ are the local displacement operators (see Sec. \ref{subsubsec:battery_formalism}). The energy gained by the battery upon charging has a much simpler form than the one in the case of the local squeezer and depends only on the parameters $|\alpha_j|$
\begin{equation}
    \Delta E_{\hat{D}_c} = \Delta E_{\hat{D}_c^1} + \Delta E_{\hat{D}_c^2} = |\alpha_1|^2 + |\alpha_2|^2.
    \label{eq:deltaE_2-mode_tau_disp}
\end{equation}
Again, similar to the previous case, we can calculate $\Delta \sigma^{(2)} = f_D^{(2)}(r,|\alpha_j|,\phi_j,\tau)$ and since $\alpha_j = \sqrt{\Delta E_{\hat{D}^{j}_{c}}}$ (from Eq. \eqref{eq:deltaE_2-mode_tau_disp}), we obtain $\Delta \sigma^{(2)}$ in terms of $r, \Delta E_{\hat{D}_c^j}, \phi_j$s and $\tau$.
By fixing a value of $r$, we again find that $\Delta E_{\hat{D}_c^1} = \Delta E_{\hat{D}_c^2} = \Delta E_{\hat{D}_c}/2$, leads to minimal fluctuation. Moreover, the optimal charging precision turns out to be independent of $\tau$ (see Fig. \ref{fig:2-3_modes}(b)). Thus for every value of $\tau$, we obtain a particular set of $(\phi_1, \phi_2)$ (e.g. for $\tau=0,~\phi_1=\pi/2 \text{ and } \phi_2=0$) which yields the same minimum value of $\Delta \sigma^{(2)}$. 

 \textcolor{black}{To estimate the work fluctuation, here we need to compute  $\hat{H}_B'=\hat{D}_c^\dagger\hat{H}_B\hat{D}_c$. By employing the relation $\hat{D}_j^\dagger(\alpha_j)\hat{a}_j\hat{D}_j(\alpha_j)=\hat{a}+\alpha$, we derive 
$$\hat{H}_B'=\sum_{j=1,2}\hat{N}_j+\alpha_j^*\hat{a}_j+\alpha_j\hat{a}_j^\dagger+|\alpha_j|^2.$$ Since the Wigner function corresponding to the state $\rho_0$ is symmetric in its arguments, the odd moments of the field operators vanish. This results in $\langle\{\hat{H}_B',\hat{H}_B\}\rangle_{\rho_0}=4(\alpha_1^2+\alpha_2^2+2\cosh 2r)\sinh^2r$. Consequently,  we obtain $\Delta W^{(2)}=g_D^{(2)}(r, \tau, \Delta E_{\hat{D}_c^j}, \phi_j)$. For a specific value of $r$, we find that the minimal work fluctuation is achieved with equal charging via each mode at optimal phase configuration. Like the charging precision, the optimal phase configuration to achieve minimal work fluctuation is also $\tau$ dependent (e.g., for $\tau=0$, $\phi_1=0$, and $\phi_2=3\pi/2$, or for $\tau=1/2$, $\phi_1+\phi_2=\pi$). However, the value of $\Delta W^{(2)}_{\min}$ is again independent of $\tau$ (see Fig. \ref{fig:2-3_modesW} (b)).}

\subsection{Entangled three-mode battery}
\label{subsec:3-mode_entangled}
Let us now move to the initial state of the QB as a three-mode entangled state, typically prepared using a tritter \cite{braunstein1998,loock2000}, which consists of two beam splitters. This class of three-mode entangled states can be described by the transmission coefficients $\tau_1$ and $\tau_2$ of the two beam splitters, $BS_{12}$ and $BS_{23}$, which operate on the mode pairs $(1,2)$ and $(2,3)$ respectively. In this setup, the first and third input modes of the tritter are squeezed vacuum in the momentum quadrature, while the second input mode is squeezed vacuum in the position quadrature, all with a squeezing strength $r$. This configuration results in the specified class of three-mode entangled states. \textcolor{black}{A detailed parameterization of the state in terms of its displacement vector and covariance matrix can be found in Appendix. \ref{app:states}.}

The energy of such an uncharged battery is \(E_0^{(3)} = 3 \sinh^2r\). Again local displacement and squeezing unitaries are applied individually on the three modes to charge the battery.
\subsubsection{Charging with local squeezing and displacement}
\label{subsubsec:3-mode_tau_charging}
When local squeezing unitaries are applied to three modes, each with a degree $\delta_j$ and angle $\theta_j$ \((j=1,2,3)\), the increase in the energies of the individual battery modes are given by
\begin{eqnarray}
   \nonumber \Delta E_{\hat{S}_c^1} =  \sinh \delta_1 \Big[ && (2 \tau_1-1) \cosh \delta_1 \cos \theta_1 \sinh 2r + \\
&& \sinh \delta_1 \cosh 2r \Big], \label{eq:3-mode_tau_deltaE1} \\
 \nonumber \Delta E_{\hat{S}_c^2} = \sinh \delta_2 \Big[ && \cosh \delta_2 \cos \theta_2 \sinh 2r (1-2 \tau_1 \tau_2)+\\
 && \sinh \delta_2 \cosh 2r \Big], \label{eq:3-mode_tau_deltaE2} \\
\nonumber \Delta E_{\hat{S}_c^3} = \frac{1}{4} e^{-2 r} \Big[ && \left(e^{4 r}-1\right) \sinh 2 \delta_3 \cos \theta_3  (2 \tau_1 (\tau_2-1)+1)  \\ 
&&    +2 \left(e^{4 r}+1\right) \sinh ^2 \delta_3 \Big],  \label{eq:3-mode_tau_deltaE3} 
\end{eqnarray}
with total energy increment, $\Delta E_{\hat{S}_c} = \sum_{j = 1}^3 \Delta E_{\hat{S}_c^j}$. Considering that we distribute the energy $\Delta E_{\hat{S}_c}$ as $k_1 \Delta E_{\hat{S}_c},  k_2 \Delta E_{\hat{S}_c}$ and $(1 - k_1 - k_2) \Delta E_{\hat{S}_c}$ in the three modes such that $k_1 + k_2 \leq 1$, we observe that \textcolor{black}{both $\Delta W^{(3)}_{\min}$ and} $\Delta \sigma^{(3)}_{\min}$ occur at $k_1 = k_2 = 1/3$, i.e., $\Delta E_{\hat{S}_c^j} = \Delta E_{\hat{S}_c}/3$. Further analysis proceeds in a manner similar to the two-mode case. 
In this scenario, the minimum charging precision is obtained, for any initial squeezing $r$ and $\Delta E_{\hat{S}_c}$, when $\theta_j = \pi/2$ and is independent of $\tau_1$ and $\tau_2$ (see Fig. \ref{fig:2-3_modes} (c)). \textcolor{black}{This implies that the genuinely entangled state, the biseparable configuration (obtained with $\tau_2 = 0$), as well as the fully separable state (obtained by setting $\tau_1 = 0$), perform equally well.} Furthermore, when the battery comprises fully separable modes, $\Delta \sigma_{\min}^{(3)}$ is independent of the squeezing angles $\theta_j$, thereby providing more freedom during the charging process. 

\begin{figure*}
    \centering
    \includegraphics[width=1.00 \linewidth]{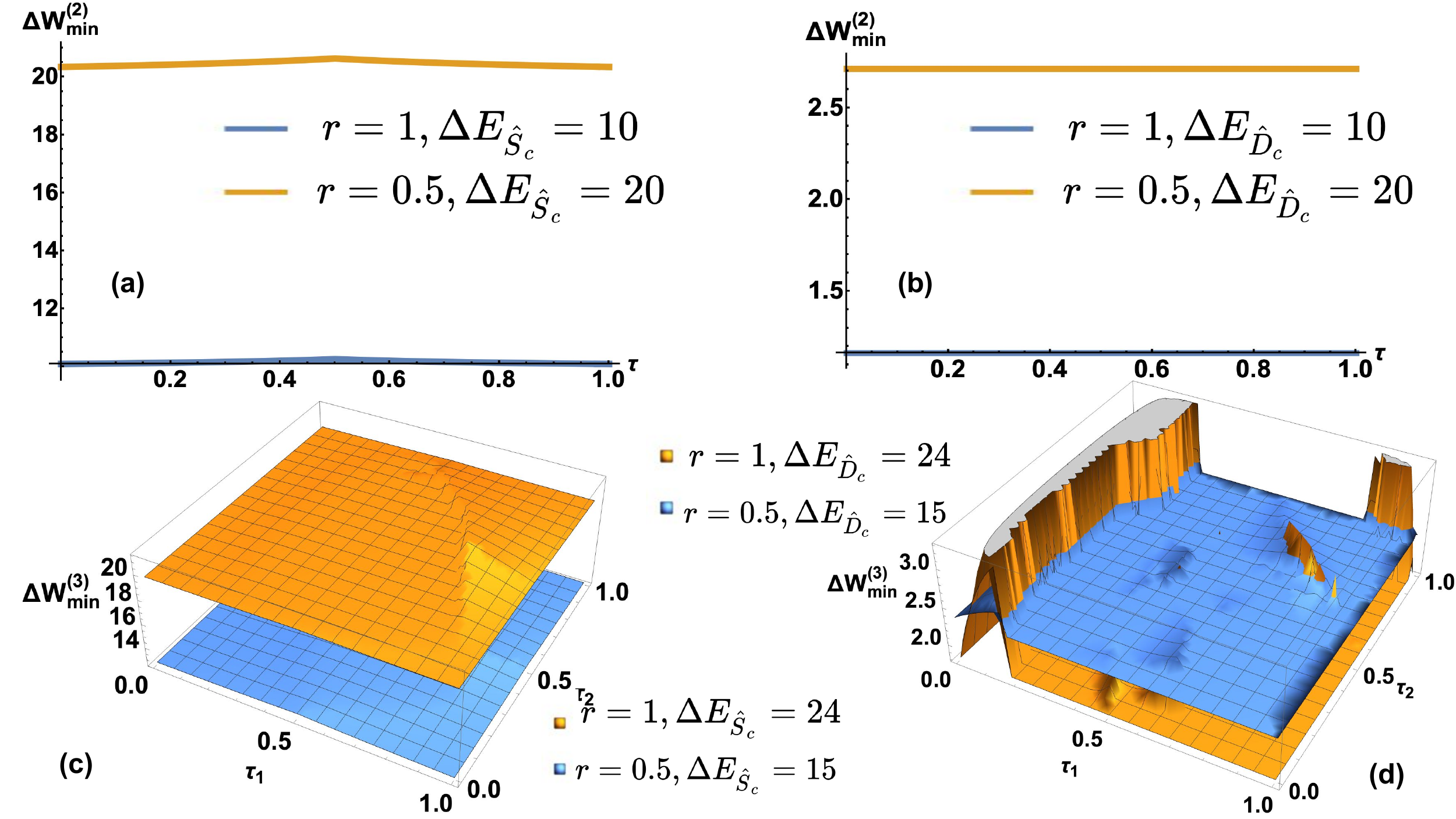}
    \caption{(Color Online.) \textbf{Behavior of the optimal work fluctuation, $\Delta W_{\min}^{(\mathcal{N})}$ for two- and three-mode entangled initial states}.  All specifications are the same as in Fig. \ref{fig:2-3_modes}. All the axes are dimensionless.}
    \label{fig:2-3_modesW}
\end{figure*}

\textcolor{black}{On the other hand, $\Delta W^{(3)}$ can be computed similarly as described in the two-mode scenario. Although, for the sake of brevity, we refrain from writing the exact expression of $\langle\{\hat{H}_B',\hat{H}_B\}\rangle_{\rho_0}$, using Eqs. \eqref{eq:3-mode_tau_deltaE1} - \eqref{eq:3-mode_tau_deltaE3}, it can be shown that we end up with $\Delta W^{(3)} = g_S^{(3)}(r,\Delta E_{\hat{S}_c^j}, \theta_j, \tau_j)$ for some function $g_S^{(3)}$. When the initial energy (determined by $r$) and total energy increment, $\left(\Delta E_{\hat{S}_c}\right)$, are held constant for a given state, the minimal work fluctuation can be attained by optimizing the phase configuration. However, the optimal phase configuration depends on $\tau_1$ and $\tau_2$, e.g., in case of $\tau_1=\tau_2=0$, optimality is achieved at $\theta_1=\pi$ and $\theta_2=\theta_3=0$, whereas, for $\tau_1=\tau_2=1$ it is $\theta_2=\pi$ and $\theta_1=\theta_3=0$. Fig. \ref{fig:2-3_modesW} (c) illustrates how $\Delta W_{\min}^{(3)}$ varies with the state parameters $\tau_1$ and $\tau_2$. It shows that the minimum value is attained at $\tau_1=0$, regardless of the value of $\tau_2$.}

In case of charging via local displacements, having parameters $\alpha_j = |\alpha_j|e^{i\phi_j}$, the increase in the energy of the battery is given as
\begin{equation}
    \Delta E_{\hat{D}_c} = \sum_{j = 1}^3 \Delta E_{\hat{D}_c^j} = \sum_{j = 1}^3 |\alpha_j|^2,
    \label{eq:3-mode_tau_deltaE}
\end{equation}
where $\Delta E_{\hat{D}_c^j}$ is the energy change in mode $j$.

Given an initial squeezing strength $r$ and an energy change $\Delta E_{\hat{D}_c}$, the charging precision attains minimum at some set of values of $\tau_j$s with some optimal choice of $\phi_j$s when all the modes are equally discharged (see Fig. \ref{fig:2-3_modes} (d)). Specifically, at $\tau_1=0$, we find that the optimal choice of phases, which yields the minimum charging precision $\Delta \sigma_{\min}^{(3)}$, is $\phi_1=\pi, \phi_2=\phi_3=\pi/2$. 
Furthermore, at $\tau_1=0$, $\Delta \sigma_{\min}^{(3)}$ is independent of $\tau_{2}$, implying that we can dispense with the beam splitters altogether, thereby dealing with three separate squeezed modes. 

\textcolor{black}{The work fluctuation, $\Delta W^{(3)}$, has a much simpler form. In particular, following the same prescription provided in the two-mode scenario, we obtain $\langle\{\hat{H}_B',\hat{H}_B\}\rangle_{\rho_0}=3\left(2\alpha_1^2+2\alpha_2^2+2\alpha_3^2+5\cosh 2r -1\right)\sinh^2 r$. 
By replacing $|\alpha_j|$ with $\sqrt{\Delta E_{\hat{D}_c^j}}$, we can write $\Delta W^{(3)} = g_D^{(3)}(r, \tau, \Delta E_{\hat{D}_c^j}, \phi_j)$. At an initial squeezing strength, $r$, and for a fixed energy increment, $\Delta E_{\hat{D}_c}$, the work fluctuation becomes minimum at some optimal choice of $\phi_j$s when equally charged through each mode. The optimal choice of $\phi_j$s depends on the state parameters, e.g., for fully separable states (i.e., $\tau_1=0$) optimal phases are $\phi_1=0,$ and $\phi_2=\phi_3=\pi/2$, whereas for maximally genuinely entangled state (i.e., $\tau_1=1/2$ and $\tau_2=1/2$),  optimality is attained at $\phi_1=\phi_2=\phi_3=3\pi/2$. Observing Fig. \ref{fig:2-3_modesW} (d), it is evident that both fully separable states ($\tau_1=0$) and certain entangled states exhibit minimal work fluctuation.} \\
The results obtained for the two- and three-mode batteries can be consolidated as follows.\\
\\
\textcolor{black}{\textbf{Observation $1$}. \textit{For two-mode and three-mode batteries charged in all the modes through local squeezing and displacement unitaries, entanglement plays no role in the optimal performance as quantified by the charging precision, $\Delta \sigma_{\min}$, and the work fluctuation $\Delta W_{\min}$.}} \\

\textcolor{black}{\textbf{Remark}. Instead of using an entangled state as the initial state, we can do the reverse operation as well, i.e., we take a two-mode separable state (\(2-\mathcal{SMSV}\), i.e., product of two $\mathcal{SMSV}$ states) as the initial state and charge the battery by a global squeezing operation where entanglement can, in principle, be generated through the evolution of the system. We show in Appendix \ref{app:ent_squeezer} that entanglement generation through time evolution does not provide any advantage over local squeezing operation. This kind of scenario may appear in any general physical system as entanglement generation is neither necessary nor sufficient criteria to achieve quantum advantage \cite{hovhannisyan2013}.}\\

\textcolor{black}{
\textbf{Observation $2$}. \textit{In the cases of both the two-mode and the three-mode batteries, charging via displacement operation is favorable over charging with local squeezing, since $\Delta \sigma_{\text{min}}$ and $\Delta W_{\text{min}}$ are much lower in the case for charging via local displacement.}}\\

\textcolor{black}{\textbf{Note.} It has been well established that entanglement plays a crucial role in the performance of quantum batteries comprising finite dimensional systems. However, the dynamics of systems comprising modes with continuous spectrum is very different from that of batteries built with discrete variable qudits. In our work, we have considered two paradigmatic Gaussian processes as charging operations viz., displacement and squeezing, which are easily realizable in experiments. Our conjecture is that entanglement in the battery or the charging operation would not prove to be more efficient even when other charging processes are used. This is because all Gaussian operations can be composed of displacements, squeezing, phase shifters (which cannot increase the energy, and are thus not considered as chargers), and beam splitters. Although beam splitters can induce entanglement between the involved modes and have not been considered in this work, the fact that displacement operations (which can never create entanglement) offer the best figures of merit, motivates us to conclude that local operations are the best method to operate a continuous variable battery. Instead of using Gaussian operations, non-Gaussian operations, e.g., photon subtraction may play an advantageous role in the case of an entangled state, but the implementation of such non-Gaussian operation is highly probabilistic. This further makes our results more intriguing. Our work suggests that using continuous variable systems as batteries as well as the usage of Gaussian unitaries is more economical since entangling operations need not be used for better performance.}


\section{$\mathcal{N}$-mode separable battery}
\label{sec:N-mode_sep}
\textcolor{black}{Observation $1$ allows us to argue that the unentangled state is the appropriate choice as an initial state, which is influenced by the results obtained in the preceding section. Analyzing the results of the previous section, we also notice the following:}  

\begin{enumerate}
    
    \item The phase parameters involved in the charging process (i.e., squeezing or displacement angles) play a crucial role in the optimization of the charging precision.
    

    \item It turns out that equal charging in all the modes corresponds to the best charging configuration.
\end{enumerate}
It is thus natural to take an $\mathcal{N}$-mode separable state with all the modes having equal initial squeezing ($r$), as the initial state of an $\mathcal{N}$-mode battery and consider the charging operations as local. The initial state in this case can be represented by the displacement vector and covariance matrix as
\begin{eqnarray}
&&\textbf{d}_{0}^{(\mathcal{N})} = (0,0,...,0)_{1\times2\mathcal{N}}^T,\\
\text{and}\nonumber
\label{eq:n-mode_jnitial_disp}\\
&&\Xi_{0}^{(\mathcal{N})} = \frac{1}{2}\text{diag}(e^{2r},e^{-2r},...,e^{2r},e^{-2r})_{2\mathcal{N}\times2\mathcal{N}}.
\label{eq:n-mode_initial_cov}
\end{eqnarray}
The initial average energy of this system is ${E_0^{(\mathcal{N})}}=\mathcal{N} \sinh ^2 r$,
while the average of the squared energy takes the form as
\begin{equation}
    {E_0^{2(\mathcal{N})}}=\frac{1}{2} \mathcal{N} \sinh ^2 r \Big(2+(\mathcal{N}+2) \cosh (2 r)-\mathcal{N}\Big).
\end{equation}
The corresponding second moment of the energy turns out to be $V({\rho_0}) = \mathcal{N} \cosh^2r \sinh^2r$. By applying local squeezing and displacement unitaries in each mode, the variation of the figures of merit of the battery, against the number of modes, can be obtained. Before presenting the results, let us state a lemma, which will be frequently applied.\\\\
\textbf{Lemma $1$}. \textit{For any multivariate function $f(x_1,x_2,\dots x_n)$ which is symmetric with respect to permutations of its arguments, the global extremum occurs at $x_1 = x_2 = \dots = x_n$}.\\\\
\textit{Proof}. Since $f$ is a symmetric function of its arguments, we have $f(x_1,x_2,\dots x_n) = f(x_2,x_1,\dots x_n) = \dots$ and all possible permutations of $x_{j}$. Suppose that $f_{\min(\max)} = f(x_1^0,x_2^0,\dots x_n^0)$. Then all possible permutations of $x_j^0$ as the argument of $f$ would still yield $f_{\min(\max)}$. Since the extremum is a global one, this implies that $x_1^0 = x_2^0 = \dots x_n^0$. $\hfill \blacksquare$\\
A direct consequence of this lemma is that the figures of merit considered here, being symmetric functions of the increment in mode energies $\Delta E_{{U_C^j}}$, can be minimized by distributing the entire energy $\Delta E_{{U_C}}$ equally in each mode i.e., by choosing $\Delta E_{U_C^j} = \Delta E_{U_C}/\mathcal{N} ~ \forall j$. This has been extensively verified for all the models under consideration.
\subsection{Charging with local squeezing}
\label{subsec:N-mode_squeezing}
The charging of an $\mathcal{N}$-mode battery by applying local squeezing in each mode is performed with the modal squeezing parameter, $\zeta_j = \delta_j e^{i \theta_j}$. Hence the total charging operator becomes $\hat{S}_c = \hat{S}_1(\zeta_1) \otimes \hat{S}_2(\zeta_2)\otimes...\otimes\hat{S}_\mathcal{N}(\zeta_\mathcal{N})$. Upon charging, the total energy is given by
\begin{equation}
     {E_{\hat{S}_c}^{(\mathcal{N})}} = \sum_{j=1}^\mathcal{N} \big(\Delta E_{\hat{S}_c^j} + \sinh^2 r\big) = \sum_{j=1}^\mathcal{N} {\langle \hat{N}_{\hat{S}_c^j} \rangle},
\end{equation}
where $\langle \hat{N}_{\hat{S}_c^j} \rangle$ \textcolor{black}{denotes the average photon number (hence, the average energy since $\omega_j=1~\forall j$)} of the mode $j$ after charging and $\Delta E_{\hat{S}_c^j}$ is the corresponding energy increment in that mode, given by
\begin{equation}
     \Delta E_{\hat{S}_c^j} = \sinh \delta_j \big(\cosh \delta_j \cos\theta_j \sinh 2 r+\sinh \delta_j \cosh 2 r\big).
\end{equation}
Hence, the total increase in energy after charging reads as 
\begin{equation}
     \Delta E_{\hat{S}_c}^{(\mathcal{N})} = \sum_{j=1}^\mathcal{N} \Delta E_{\hat{S}_c^j} = {E_{\hat{S}_c}^{(\mathcal{N})}} - {E_0^{(\mathcal{N})}}.
\end{equation}
On the other hand, the average of the squared energy for the $\mathcal{N}$-mode system is given by
\begin{equation}
    E_{\hat{S}_c}^{2(\mathcal{N})}=\langle \hat{N}_{\hat{S}_c}^{2(\mathcal{N})} \rangle =\sum_{j = 1}^\mathcal{N} \langle \hat{N}_{\hat{S}_c^j}^2 \rangle + 2\sum_{j = 1}^{\mathcal{N} - 1} \sum_{k>j}^\mathcal{N} \langle \hat{N}_{\hat{S}_c^j} \rangle \langle \hat{N}_{\hat{S}_c^k} \rangle,
\end{equation}
with $\langle \hat{N}_{\hat{S}_c^j}^2 \rangle$ being the mean of squared energy of the  mode $j$, which takes the form as 
\begin{eqnarray*}
\langle \hat{N}_{\hat{S}_c^j}^2 \rangle&&=\frac{1}{32} \Big(3 \cosh 4 \delta_j +\\&& 4 \sinh 2 \delta_j  \sinh 2 r (3 \sinh 2 \delta_j  \cos 2 \theta_j  \sinh 2 r-4 \cos \theta_j )+\\&&12 \sinh 4 \delta_j  \cos \theta_j  \sinh 4 r-16 \cosh 2 \delta_j  \cosh 2 r+\\&&(9 \cosh 4 \delta_j +3) \cosh 4 r+1\Big).
\end{eqnarray*}
Having the expressions for the averages of the energy and the squared energy, one can find the standard deviation in the energy before and after the charging process. Thereafter, from Eqs. \eqref{eq:sigma_defn} and  \eqref{eq:battery_V}, we can find the charging precision $\Delta\sigma_{\hat{S}_c}^{(\mathcal{N})}$ as a function of charging parameters, i.e., $\{\delta_j ,\theta_j\}_{j=1}^\mathcal{N}$ and the initial squeezing strength $r$ of the $\mathcal{N}$-mode battery. \\
Note that we can also write $\Delta\sigma_{\hat{S}_c}^{(\mathcal{N})}$ in terms of $\{\Delta E_{\hat{S}_c^j} ,\theta_j\}_{j=1}^\mathcal{N}$ and $r$ by replacing $\delta_{j}$ as
\begin{widetext}
\begin{equation}
   \delta_{j} = \frac{1}{2}\Big[\ln \left(\frac{\sqrt{4 \Delta E_{\hat{S}_c^j} (\Delta E_{\hat{S}_c^j}+\cosh 2 r)+\cos ^2\theta_j  \sinh ^2 2 r}+2 \Delta E_{\hat{S}_c^j}+\cosh 2 r}{\cos \theta_j  \sinh 2 r+\cosh 2 r}\right)\Big].
   \label{eq:delta_N-mode}
\end{equation}
\end{widetext}
In order to calculate the work fluctuation, we first need to find $\expval{\{\hat{H}_B,\hat{H'}_B\}}_{\rho_0}$ with respect to the initial state of the battery $\rho_0$. Using some operator algebra, we can obtain the expectation value of the above-mentioned anticommutator as
\begin{widetext}
\begin{eqnarray}
\expval{\{\hat{H}_B,\hat{H'}_B\}}_{\rho_0} = \sum_{j=1}^\mathcal{N}\Big[ &&\frac{2}{\mathcal{N}} {(E_0^{2(\mathcal{N})})} \cosh 2\delta_j ~ + ~ 2 {E_0^{(\mathcal{N})}} \sinh^2 \delta_j ~ + ~ \frac{1}{2} \cos\theta_j \sinh 2r (3 \cosh 2r -1) \sinh 2\delta_j ~ + ~\nonumber\\&& \frac{\mathcal{N}-1}{\mathcal{N}} {E_0^{(\mathcal{N})}} \sinh 2r \cos\theta_j \sinh 2\delta_j \Big],
\end{eqnarray}
\end{widetext}
\begin{figure*}
    \centering
    \includegraphics[width = 1.00\linewidth]{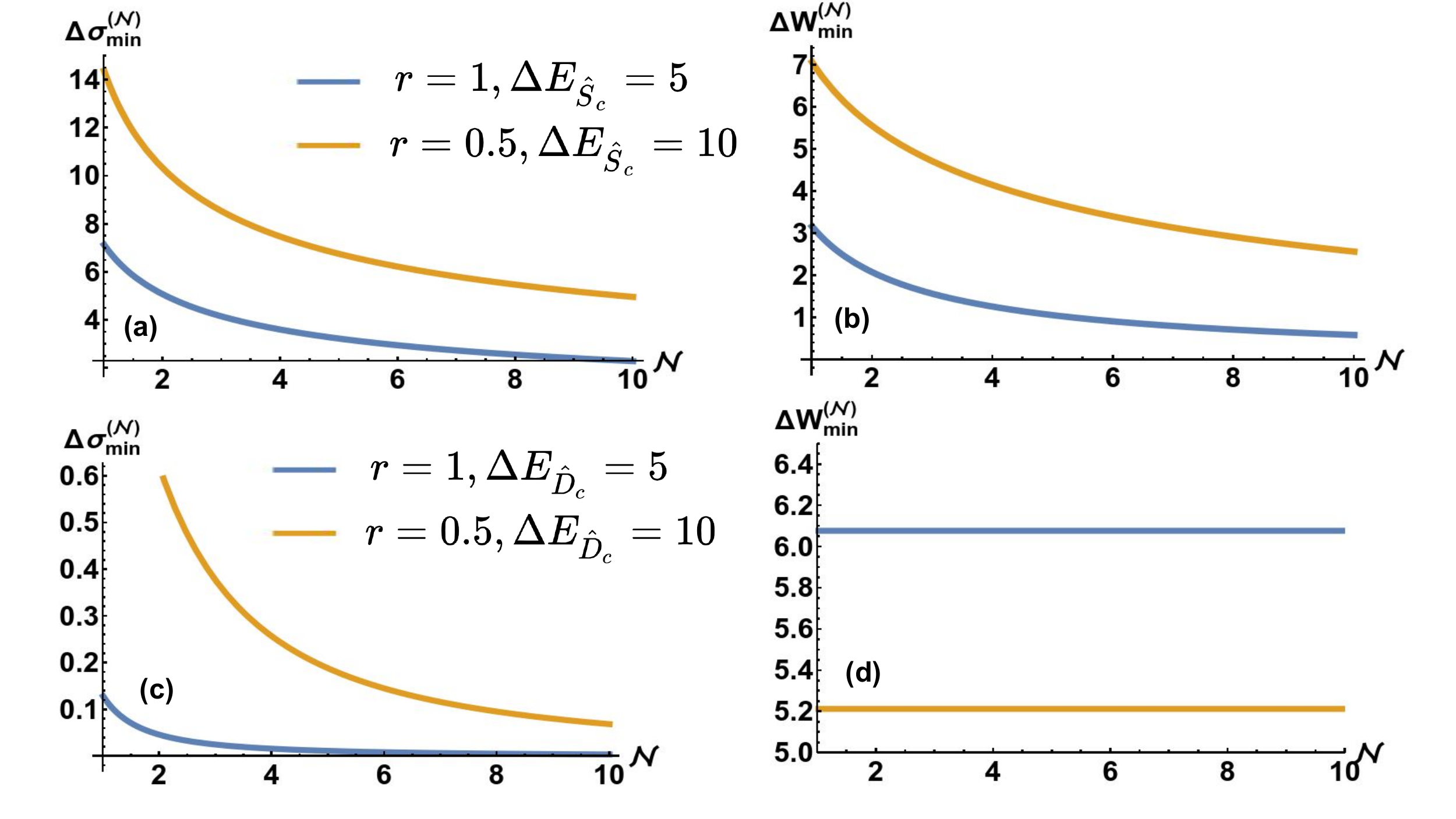}
    \caption{(Color Online.) {\bf Scaling analysis. } (a) and (c). Variation of the minimum  charging precision, $\Delta \sigma_{\min}^{(\mathcal{N})}$  (ordinate) against the number of modes, \(\mathcal{N}\). (b) and (d). The work fluctuation $\Delta W_{\min}^{(\mathcal{N})}$ (ordinate) vs $\mathcal{N}$ (abscissa). When local squeezing unitaries in (a) and (b) act as chargers, we choose $r = 1, \, \Delta E_{\hat{S}_c} = 5$ (dark blue) and $r = 1, \, \Delta E_{\hat{S}_c} = 10$ (light orange). The charging process is taken to be local displacement operations in (c) and (d). The initial squeezing $r$ and the energy increment $\Delta E_{\hat{D}_c}$ are considered as $r = 1, \, \Delta E_{\hat{D}_c} = 5$ (dark blue) and $r = 0.5, \, \Delta E_{\hat{D}_c} = 10$ (light orange).   
    All the axes are dimensionless.}
    \label{fig:sigma_w}
\end{figure*}
and using Eqs. \eqref{eq:W_defn} and \eqref{eq:delta_N-mode}, we can obtain the expression of $\Delta W_{\hat{S}_c}^{(\mathcal{N})}$ in terms of $\{\Delta E_{\hat{S}_c^j} ,\theta_j\}_{j=1}^\mathcal{N}$ and $r$. \\
\subsubsection*{Advantage of multiple modes in the battery for charging with local squeezing}
According to Lemma $1$, the minimum for both the figures of merit are obtained when each mode has its energy raised by $\Delta E_{U_C^j}=\Delta E_{U_C}/\mathcal{N}$. An interesting point to note is that for a fixed initial squeezing $r$ of the battery modes, an equal energy increment in each mode, implies that all the modes have equal charging parameters. In other words, we have $\delta_1 = \delta_2 = \dots = \delta_\mathcal{N} = \delta$ and $\theta_1 = \theta_2 = \dots = \theta_\mathcal{N} = \theta$. This simplifies our analysis considerably since we only need to optimize over a single parameter, i.e., $\theta$, in this charging scenario.\\\\
\emph{Charging precision:} In the optimal charging configuration (i.e., $\Delta E_{\hat{S}_c^j} = \Delta E_{\hat{S}_c}^{(\mathcal{N})}/\mathcal{N}$), we see that there is no phase dependence in the $\Delta\sigma^{(\mathcal{N})}$. 
In particular, we have the following observation:

\textbf{Observation 3}. \textit{For an $\mathcal{N}$-mode separable battery undergoing equal charging in all the modes via squeezing operations, the charging precision $\Delta \sigma^{(\mathcal{N})}$ is independent of the squeezing angles $\theta_j$}.\\\\
On taking the derivative of $\Delta \sigma^{(\mathcal{N})}$ with respect to $\theta_j$, it is observed that $\pdv{\Delta \sigma^{(\mathcal{N})} (r, \Delta E_{\hat{S}_c^j}, \theta_j)}{\theta_j} = 0 ~\forall j$ when $\Delta E_{\hat{S}_c^j} = \Delta E_{\hat{S}_c}/\mathcal{N}$. 

The minimum charging precision ($\Delta\sigma_{{\min}}^{(\mathcal{N})}$) as a function of the total increase in energy of the battery $\Delta E_{\hat{S}_c}^{(\mathcal{N})}$ is depicted  against the number of modes $\mathcal{N}$ in Fig. \ref{fig:sigma_w} (a). Clearly, $\Delta \sigma^{(\mathcal{N})}_{{\min}}$ decreases with $\mathcal{N}$. Precisely, from Fig. \ref{fig:sigma_w} (a), we can see that $\Delta\sigma_{{\min}}^{(\mathcal{N})}$ is roughly proportional to $\mathcal{N}^{-0.49}$ which indicates that there exists an advantage with the increase in the number of modes which scales as $\sim \frac{1}{\sqrt{\mathcal{N}}}$.\\

\emph{Work fluctuation:} In case of $\Delta W^{(\mathcal{N})}$, unlike the charging precision, we note that $\Delta W^{(\mathcal{N})}$ is phase dependent in the optimal charging configuration. The minimum of work fluctuation, $\Delta W_{{\min}}^{(\mathcal{N})}$, is achieved when all the squeezing angles vanish. In Fig. \ref{fig:sigma_w} (b), we plot $\Delta W_{{\min}}^{(\mathcal{N})}$ with $\mathcal{N}$ while keeping $\Delta E_{\hat{S}_c}^{(\mathcal{N})}$ fixed at some values. We demonstrate that with an increase in the number of modes, the optimal work fluctuation decreases, and hence the efficiency of the battery enhances. We can also see that $\Delta W_{{\min}}^{(\mathcal{N})}$ can be approximated by a function proportional to $\mathcal{N}^{-0.49}$ which matches exactly with the charging precision. \\ \\

\subsection{Local displacement as charging operation}
\label{subsec:N-mode_disp}

Let us now move to a situation where charging is achieved via local displacements in each mode. In this case, the entire charging unitary can be represented as $\hat{D} = \otimes_{j = 1}^\mathcal{N} \hat{D}_j (\alpha_j)$, where $\alpha_j = |\alpha_j| e^{i \phi_j}$ as defined before. In this case, the change in energy of the battery upon charging is given by \(\Delta E_{\hat{D}_c} = \sum_{j = 1}^{\mathcal{N}} |\alpha_j|^2\)
where $\Delta E_{\hat{D}_c^j} = |\alpha_j|^2$ is the energy gained by the battery mode, $j$. The average energy and the mean squared energy for the $j$-th mode are expressed as
\begin{eqnarray}
  \langle \hat{N}_{\hat{D}_c^j} \rangle =   && \sinh^2r + |\alpha_j|^2  \label{eq:N_i_N-mode}\\
   \nonumber \langle \hat{N}_{\hat{D}_c^j}^2 \rangle =   && \alpha_j ^2(\alpha_j ^2 - 1 + \cos 2 \phi  \sinh 2 r)+\left(2 \alpha_j ^2-\frac{1}{2}\right) \cosh 2 r \\
     && +\frac{1}{8} (3\cosh 4 r + 1). \label{N2_N-mode}
\end{eqnarray}
Therefore, the average energy and the mean squared energy for the $\mathcal{N}$-mode system can be computed as
\begin{eqnarray}
    {E_{\hat{D}_c}^{(\mathcal{N})}} = && \sum_{j = 1}^\mathcal{N} \langle \hat{N}_{\hat{D}_c^j} \rangle, \label{eq:E1_disp_N} \\
     E_{\hat{D}_c}^{2(\mathcal{N})} = && \sum_{j = 1}^\mathcal{N} \langle \hat{N}_{\hat{D}_c^j}^2 \rangle + 2\sum_{j = 1}^{\mathcal{N} - 1} \sum_{l>j}^\mathcal{N} \langle \hat{N}_{\hat{D}_c^j} \rangle \langle \hat{N}_{\hat{D}_c^l} \rangle. \label{eq:E2_disp_N}
\end{eqnarray}
Consequently, the variance of energy in the state after the displacement charging operation, $V({\rho_1})$, can be determined using Eq. \eqref{eq:battery_V}. Further, as described in Eq. \eqref{eq:sigma_defn}, we can calculate the charging precision as $\Delta \sigma^{(\mathcal{N})} = f_D^{(\mathcal{N})}(r, \Delta E_{\hat{D}_c^j}, \phi_j)$ for some function $f_D^{(\mathcal{N})}$.

The expression for $\Delta W^{(\mathcal{N})}$ can also be computed by noting that, for charging via displacement, we have
\begin{equation}
    \expval{\{\hat{H'}_B,\hat{H}_B\}}_{\rho_0} = 2 \sum_{j = 1}^\mathcal{N} |\alpha_j|^2  E_0^{(\mathcal{N})}  + 2  E_0^{2(\mathcal{N})} ,
    \label{eq:N-mode_disp_anticomm}
\end{equation}
from which we can obtain the work fluctuation using Eq. \eqref{eq:W_defn}.
\subsubsection*{Multimodal advantage for charging with local displacement}
Akin to the case of charging via local squeezing, here too an equal charging in all the modes, i.e., $|\alpha_j|=|\alpha|$ and $\phi_j=\phi ~ \forall ~ j$, corresponds to the optimal charging configuration and yields the best figures of merit according to Lemma $1$. In this situation, we find that both $\Delta \sigma_{\text{min}}^{({\mathcal{N}})}$ as well as $\Delta W_{\text{min}}^{(\mathcal{N})}$, are determined by \textcolor{black}{the phase factors of the charging unitaries, $\phi$.}\\\\
\emph{Charging precision: }At $\phi = \pi/2$, we obtain that the least charging precision steadily decreases with $\mathcal{N}$ for a fixed $r$ and $\Delta E_{\hat{D}_c}^{(\mathcal{N})}$. Interestingly, the scaling with $\mathcal{N}$ is found to $\frac{1}{\mathcal{N}^{3/2}}$ (see Fig. \ref{fig:sigma_w} (c)). 

{\bf Observation $4$}. \textit{The displacement unitary can provide better stability of the quantum battery with the increase of the number of modes, in terms of the charging precision, in comparison with the squeezing one.} \\\\

\emph{Work fluctuation: } The work fluctuation also attains its least value when we fix $\phi = \pi/2$. However, our numerical studies show that $\Delta W^{(\mathcal{N})}$ is independent of the number of modes, and in the optimal charging configuration, it remains the same for all $\mathcal{N}$ as shown in Fig. \ref{fig:sigma_w} (d) at a fixed value of the initial squeezing $r$ and the energy change $\Delta E_{\hat{D}_c}^{(\mathcal{N})}$.\\

Thus an increase in the number of modes has no effect on the work fluctuation for charging via local displacement operations, although the multimode advantage is evident from the behavior of $\Delta \sigma^{(\mathcal{N})}$.

\section{Conclusion}
\label{sec:conclu}

In recent years, different designs of quantum batteries have been proposed either to increase the storage capacity and stabilize their energy extraction or to take care of their implementations in certain physical systems. Till now, the majority of works on quantum batteries deal with discrete spin systems, although CV quantum systems can be a potential candidate for energy storage devices. 
In the implementation of quantum information processing tasks, such systems have been demonstrated to have some advantages over finite dimensional systems, including the stability of QBs.
In CV systems, the fluctuation in the extraction (storage) of energy from (in) the QB during discharging (charging) can also be quantified.

We considered \textcolor{black}{a class of pure multimode Gaussian states as the initial state of the CV quantum battery, in which tuning of the parameters leads to both entangled and separable states.}  To charge the battery, local squeezing and displacement operations as well as entangling squeezing operations were carried out.  
 We established that increasing the number of modes contributes positively to the stability of the system by conducting the whole analysis in the phase space description and taking into account the second moments in the change of energy.
One of the counterintuitive results was that after optimizing the relevant parameters, entanglement between the modes did not turn out to be a prerequisite for creating a stable quantum battery. 
In sharp contrast, multimodal entanglement is required for quantum advantage in quantum communication protocols including cryptography and metrological problems. 
 However, our results immediately imply that such a QB is economically less expensive to create than a QB requiring entanglement since separable multimode states turn out to be the optimal ones with local unitary operations, especially displacements.
 
Furthermore, we established that while the strengths of the squeezing and displacement operations of the charger determine the energy to be stored in the battery, the behavior of the considered figures of merit, namely the optimal charging precision and the work fluctuation, is purely dictated by the corresponding phases. It demonstrates that, in contrast to standard CV protocols like dense coding, teleportation, and quantum illumination, which primarily rely on the magnitude of the squeezing or displacement present in the resource state, QB requires other features for good stability. This possibly indicates that quantum communication or quantum sensing protocols do not require the same resources as thermodynamic tasks like storing energy in batteries. 
Moreover, the modal advantage in the stability of the battery provides the theoretical groundwork for developing effective thermal storage devices in the future.

\section*{Acknowledgment}
\label{sec:Ack}
We acknowledge the support from the Interdisciplinary Cyber-Physical Systems (ICPS) program of the Department of Science and Technology (DST), India, Grant No.: DST/ICPS/QuST/Theme- 1/2019/23. This research was supported in part by the `INFOSYS scholarship for senior students'.

\appendix
 \section{Continuous variable (CV) formalism} 
 \label{sec:CV}
 As the name suggests, CV systems are those whose relevant degrees of freedom admit a continuous spectrum. We need an infinite dimensional Hilbert space to describe the system.The Hamiltonian description of the energy of an $\mathcal{N}$-mode system can be represented as
\begin{equation}
	\hat{H}=\sum_{j=1}^{\mathcal{N}}\hat{H}_j \quad \text{where} \quad \hat{H}_j =\hbar\omega_j\left ( \hat{a}_j^{\dagger}\hat{a}_j +\frac{1}{2}\right ),
 \label{eq:CV_hamiltonian}
\end{equation}
with \(\omega_j\) being the frequency of the mode $j$. Here \(\hat{a}_j\) (\(\hat{a}_j^\dagger\)) is the annihilation (creation) operator \textcolor{black}{(together, field operators)} for the corresponding mode and they follow the bosonic commutation rule, $[\hat{a}_j, \hat{a}_j^\dagger] = 1$. In terms of position operators, \(\hat{x}_j\) and momentum operators, \(\hat{p}_j\) \textcolor{black}{(together, quadrature operators)}, we have
	\begin{equation}
		\hat{x}_j=\frac{\hat{a}_j^{\dagger}+\hat{a_j}}{\sqrt{2}}, \quad \text{and} \quad \hat{p}_j=\frac{\hat{a_j}-\hat{a}_j^{\dagger}}{i\sqrt{2}} .
  \label{eq:CV_xp}
	\end{equation}
	Including all the modes, we can group together the canonical operators in the vector form
	\begin{equation}
		\hat{R}=\left (\hat{x}_1,\hat{p}_1,\ldots \hat{x}_{\mathcal{N}},\hat{p}_\mathcal{N}\right )^T,
\label{eq:CV_R}
	\end{equation}
	such that the commutation rules can be rewritten as
	\begin{equation}
		\left[\hat{R}_k,\hat{R}_l^{\dagger}\right]=i \mathcal{M}_{kl}\quad \text{with} ~~  \mathcal{M} = \bigoplus\limits_{j=1}^{\mathcal{N}} \Omega_j,
  \label{eq:CV_commutation}
  \end{equation}
 where $i = \sqrt{-1}$. Here, $\mathcal{M}$ is the $\mathcal{N}$-mode symplectic form, and $\Omega_j$ is given by
  \begin{equation}
      \quad \Omega_j =\begin{pmatrix}
			0 & 1\\
			-1 & 0 
		\end{pmatrix}.
  \label{eq:CV_omega}
	\end{equation}
Now, we are interested in Gaussian states $\rho$ which are all the ground and thermal states of the second-order (quadratic) Hamiltonian. Such states can be fully described by their first and second moments known as the displacement vector $\mathbf{d}$ and covariance matrix \(\Xi\) in the phase space, given by 
	\begin{equation}
		d_k=\expval{\hat{R}_k}_{\rho},
  \label{eq:CV_disp}
	\end{equation}
	and
		\begin{equation}
		\Xi_{kl}=\frac{1}{2}\expval{\hat{R}_k\hat{R}_l+\hat{R}_l\hat{R}_k}_{\rho}-\expval{\hat{R}_k}_{\rho}\expval{\hat{R}_l}_{\rho},
  \label{eq:CV_cov}
	\end{equation}
	where \(\Xi\) is a real, symmetric, and positive definite matrix. Its elements are the two-point correlation functions between the \(2\mathcal{N}\) canonical variables. The displacement vector and the covariance matrix can also be redefined in terms of the moments of the creation and annihilation operators of the different modes, as was discussed in Ref. \cite{Vallone_PRA_2019}. This greatly simplifies the calculation of the means and variances of $\hat{N}_j$, which will be necessary for our analytical calculations. Note that there exists a mapping between the  two parameterizations of $d_k$ and $\Xi_{kl}$. In the phase space formalism, the evolution  operator of the state can be represented by a symplectic matrix, \(\mathcal{S}\) in terms of which we can derive the first and second moments of the evolved state as 
	\begin{eqnarray}
		\nonumber&&\mathbf{d}\rightarrow\mathbf{d}'=\mathcal{S}\mathbf{d},
  \label{eq:CV_d_transform}\\
  &&\Xi\rightarrow\Xi'=\mathcal{S}\Xi\mathcal{S}^T.
  \label{eq:CV_sigma_transform}
	\end{eqnarray}
\textcolor{black}{Alternatively, in the phase space formalism of CV systems, the states can equivalently be characterized by the Wigner function which accounts for the quasi probability distribution of the quadrature variables. For an $\mathcal{N}$-mode Gaussian state with displacement vector $\mathbf{d}$ and covariance matrix $\Xi$, the Wigner function can be represented as 
\begin{equation}
    W(\mathbf{R})=\frac{\exp\left[-\frac{1}{2}(\mathbf{R}-\mathbf{d})^T\Xi^{-1}(\mathbf{R}-\mathbf{d})\right]}{(2\pi)^\mathcal{N}\sqrt{\det(\Xi)}}.
    \label{eq:Wigner_function_Gaussian}
\end{equation}
Notice that the Wigner function corresponding to a valid state is normalized, i.e., $\int_{\mathbb{R}^{2\mathcal{N}}}W(\mathbf{R})\mathbf{dR}=1$. For a symmetrically ordered function of the field operators, $\hat{\mathcal{O}}=f(\hat{a}_j,\hat{a}_j^\dagger)$, one has 
\begin{equation}
    \tr\left[\rho\hat{\mathcal{O}}\right]=\int_{\mathbb{R}^{2\mathcal{N}}}W_\rho(\mathbf{R})\tilde{f}(\mathbf{R})\mathbf{dR},
    \label{eq:expectation_value_Wigner}
\end{equation}
where $\tilde{f}(\mathbf{R})=f(R_j+i R_{j+1},R_j-i R_{j+1})$ with $j=1,...,\mathcal{N}$.\\
For a given Gaussian state (i.e., $\mathbf{d}$ and $\Xi$ are specified), Eq. (\ref{eq:Wigner_function_Gaussian}) and (\ref{eq:expectation_value_Wigner}) can be used to find the moments of the field operators, i.e., $\langle \hat{a}^m\hat{a}^{\dagger n}\rangle$ for that state.
}

\section{Class of two- and three-mode entangled states}
\label{app:states}
\textcolor{black}{
A generic two-mode state belonging to the one-parameter family with varying entanglement content, as discussed in Sec. \ref{subsec:2-mode-entangled}, may be conveniently represented in the phase space picture by the following displacement vector and covariance matrix :
\begin{eqnarray}
&&\textbf{d}_{0}^{(2)} = \left(
\begin{array}{cccc}
 0 \\
 0  \\
 0 \\
 0  \\
\end{array}
\right), \quad \label{eq:2-mode_gen_disp} \Xi_0^{(2)} = \nonumber \left(
\begin{array}{cccc}
 \mathcal{A} & 0 & \mathcal{C} & 0 \\
 0 & \mathcal{B} & 0 & -\mathcal{C} \\
 \mathcal{C} & 0 & \mathcal{B} & 0 \\
 0 & -\mathcal{C} & 0 & \mathcal{A} \\
\end{array}
\right), \label{2-mode_gen_cov}\nonumber \\
\end{eqnarray}
where
\begin{eqnarray}
    && \nonumber \mathcal{A} = \frac{1}{2} \left(e^{-2 r}+2 \tau \sinh 2 r\right), \\
    && \nonumber \mathcal{B} = \frac{1}{2} \left(e^{2 r}-2 \tau \sinh 2 r\right), \\
    && \mathcal{C} = \sqrt{\tau (1-\tau)} \sinh 2 r.
    \label{eq:2-mode_cov_param}
\end{eqnarray}}
\textcolor{black}{Note that at $\tau=1/2$, we obtain the well known two-mode squeezed vacuum (TMSV) state.}

\textcolor{black}{On the other hand, the state used in Sec. \ref{subsec:3-mode_entangled} can be characterized by \cite{Patra_arxiv_2022}
\begin{eqnarray}
&&\textbf{d}_{0}^{(3)} = (0,0,0,0,0,0)^T, \label{eq:3-mode_gen_disp}\\
&&\text{and}
\quad\Xi_0^{(3)} = \left(
\begin{array}{cccccc}
 \mathcal{A} & 0 & \mathcal{R} & 0 & \mathcal{T} & 0 \\
 0 & \mathcal{B} & 0 & -\mathcal{R} & 0 & -\mathcal{T} \\
 \mathcal{R} & 0 & \mathcal{C} & 0 & -\mathcal{S} & 0 \\
 0 & -\mathcal{R} & 0 & \mathcal{D} & 0 & \mathcal{S} \\
 \mathcal{T} & 0 & -\mathcal{S} & 0 & \mathcal{E} & 0 \\
 0 & -\mathcal{T} & 0 & \mathcal{S} & 0 & \mathcal{F} \\
\end{array}
\right),
\label{eq:3-mode_gen_cov}
\end{eqnarray}
where 
\begin{eqnarray}
\nonumber &&\mathcal{A} = \frac{1}{2} e^{-2 r} [\left( e^{4 r}-1\right) \tau_1 +1], \label{eq:sigma0_A}\\
\nonumber &&\mathcal{B} = \frac{1}{2}\left( e^{-2 r} \tau_1 + e^{2 r} (1-\tau_1)\right), \label{eq:sigma0_B} \\
\nonumber &&\mathcal{C} = \frac{1}{2} \left(\sinh 2 r  (1 -2 \tau_1 \tau_2)+ \cosh 2 r\right), \label{eq:sigma0_C} \\
\nonumber &&\mathcal{D} = \frac{1}{2} e^{-2 r} [\left( e^{4 r}-1\right) \tau_1 \tau_2+1], \label{eq:sigma0_D} ~~\\
\nonumber &&\mathcal{E} = \frac{1}{2} \left( \sinh 2 r (1-2 \tau_1 (1-\tau_2))+ \cosh 2 r\right), \label{eq:sigma0_E}\\
\nonumber && \mathcal{F} =\frac{1}{2} e^{-2 r} [1+\tau_1 (e^{4 r} -1 )(1 - \tau_2)],  \\
\nonumber && \mathcal{R} = \sqrt{ \tau_1 \tau_2 (1-\tau_1)} ~\sinh 2 r , \label{eq:sigma0_R} \\
\nonumber && \mathcal{S} = \tau_1 \sqrt{\tau_2 (1-\tau_2) } ~\sinh 2 r , \label{eq:sigma0_S} \\
&& \mathcal{T} = \sqrt{\tau_1 (1-\tau_1) (1-\tau_2) } ~\sinh 2 r.
\label{eq:3-mode_cov_param}
\end{eqnarray}
Note that by setting $\tau_1 = 1/3$ and $\tau_2 = 1/2$, we can prepare the Basset-Hound state \cite{loock2000,adesso_1_2007,adesso_2_2007}.}

 \section{Charging through entangling squeezer}
 \label{app:ent_squeezer}

 Let us investigate the reverse scenario where the initial state is a two-mode separable state (\(2\text{-}\mathcal{SMSV}\)) while  the global squeezing operation is applied in both the modes together to charge the battery and study the role of entanglement generation in the battery. Mathematically, the two-mode product state, in the phase space, is expressed as
	
	\begin{equation}
		\Xi_{2\text{-}\mathcal{SMSV}}=\Xi_1 \oplus \Xi_2,
		\label{eq:two_smsv}
	\end{equation}
where $\Xi_j$ denotes the covariance matrix corresponding to a single-mode squeezed vacuum $(\mathcal{SMSV})$, 
	\begin{equation}
		\Xi_1=\begin{pmatrix}
			e^{2r} & 0\\
			0 & e^{-2r} 
		\end{pmatrix}=\Xi_2.
	\end{equation}
In the phase space, the symplectic matrix representation of the global squeezing operation \cite{ferraro2005} reads as
 	\begin{equation}
 		\hat{\mathcal{S}}_c=\left(
 		\begin{array}{cccc}
 			\mathcal{C} & 0 & \mathcal{A} & \mathcal{B} \\
 			0 & \mathcal{C}& \mathcal{B} & -\mathcal{A} \\
 			\mathcal{A}  & \mathcal{B} & \mathcal{C}& 0 \\
 			\mathcal{B} & -\mathcal{A}  & 0 & \mathcal{C} \\
 		\end{array}
 		\right),
 	\end{equation}
	where
	\begin{eqnarray*}
		\nonumber&&\mathcal{A}=\cos\theta\sinh \delta,\nonumber\\&&\mathcal{B}=\sin\theta\sinh \delta, \, \, \mbox{and} \\&&\mathcal{C}=\cosh \delta,
	\end{eqnarray*}
 with \(\delta\) and \(\theta\) being the squeezing strength and the squeezing angle respectively. Here, the change in energy, \(\Delta E_{\hat{S}_c}\) is a function of \(\delta,\theta,r\), i.e., \(\Delta E_{\hat{S}_c}=f(\delta,\theta,r)\). In this case, storing energy in a single mode is not possible due to the global charging operation, although we can quantify how much energy is incorporated in each mode. We find  that \(\Delta E_{{\hat{S}}_{c}}=2 \sinh^2 \delta \cosh 2r\) and \(\Delta E_{\hat{S}_{c}^1}=\Delta E_{\hat{S}_{c}^{2}}=\Delta E_{\hat{S}_c}/2\), i.e., the  total energy increment is divided between two modes equally. We proceed to check the behavior of \(\Delta\sigma^{(2)}\) with the variation of total energy  \(\Delta E_{\hat{S}_c}\) to find if there is any advantage due to quantum correlations developed while storing the energy. \textcolor{black}{By optimizing \(\Delta\sigma^{(2)}\) with respect to the charging parameter \(\theta\), we obtain \(\Delta\sigma_{\min}^{(2)}\) at $\theta = \pi/2$.} However, when the energy increments, $\Delta E_{\hat{S}_{c}^i}$ and the initial squeezing strength, $r$, are taken to be the same as in the case of charging the entangled state with local squeezing, this minimum value exceeds the value obtained in the latter scenario. It implies that the stability of the QB obtained via local operations is higher than that via global ones. Therefore, entangling operations cannot 
    enhance the stability of the battery than that obtained from the separable initial states with local chargers. 

\bibliographystyle{apsrev4-1}
\bibliography{reference1}

\end{document}